\documentclass[aps,prd,amsmath,floats,floatfix, twocolumn,
superscriptaddress,nofootinbib,showpacs]{revtex4-2}

\usepackage[T1]{fontenc}
\usepackage[utf8]{inputenc}
\usepackage{lmodern}
\usepackage{verbatim}

\usepackage[dvipsnames, usenames]{xcolor}
\definecolor{linkcolor}{rgb}{0.0,0.3,0.5}
\usepackage[hypertexnames=false, unicode, colorlinks=true, linkcolor=linkcolor,
citecolor=linkcolor, filecolor=linkcolor,urlcolor=linkcolor,
pdfusetitle]{hyperref}
\usepackage{orcidlink}

\usepackage[all]{hypcap}
\usepackage{graphicx}
\usepackage{xspace}
\usepackage{amssymb}
\usepackage{bm} % boldmath
\usepackage{enumitem,amssymb}

\usepackage{microtype}

\usepackage[english]{babel}
\usepackage{blindtext}
\usepackage{graphicx}
\usepackage{hyperref}
\usepackage{physics}
\usepackage{siunitx}
\usepackage{bm}
\usepackage{mathtools}

\graphicspath{%
  {figs/}%
}

\DeclareMathAlphabet{\mathpzc}{OT1}{pzc}{m}{it}

\begin{document}

\title{Projected Sensitivity of a Differential Torsion Sensor \\for Tests of the Weak Equivalence Principle}

\author{Jiayi~Qin\,\orcidlink{0000-0002-7120-9026}}
\email{jiayiqinphysics@gmail.com}
\affiliation{OzGrav-ANU, Centre for Gravitational Astrophysics, College of Science and Medicine, The Australian National University, Australian Capital Territory 2601, Australia.}
\affiliation{Department of Quantum Science and Technology, Research School of Physics, The Australian National University, Australian Capital Territory 2601, Australia.}

\author{Sheon~S.~Y.~Chua\,\orcidlink{0000-0001-8026-7597}}
\affiliation{OzGrav-ANU, Centre for Gravitational Astrophysics, College of Science and Medicine, The Australian National University, Australian Capital Territory 2601, Australia.}

\author{Ling~Sun\,\orcidlink{0000-0001-7959-892X}}
\affiliation{OzGrav-ANU, Centre for Gravitational Astrophysics, College of Science and Medicine, The Australian National University, Australian Capital Territory 2601, Australia.}

\author{Bram~J.~J.~Slagmolen\orcidlink{0000-0002-2471-3828}}
\affiliation{OzGrav-ANU, Centre for Gravitational Astrophysics, College of Science and Medicine, The Australian National University, Australian Capital Territory 2601, Australia.}

\hypersetup{pdfauthor={Qin et al.}}

\date{\today}

\begin{abstract}
Tests of the weak equivalence principle (WEP) are among the most sensitive probes of new long-range, composition-dependent interactions beyond general relativity and the Standard Model. We evaluate the potential of a differential torsion sensor, which combines long mechanical coherence times with optical-cavity readout and common-mode rejection, for laboratory WEP tests and fifth-force searches. For a representative configuration based on an existing prototype, with Be--Ti test masses and 180 days of coherent integration, the projected noise-limited sensitivity to the E\"otv\"os parameter is $\eta\simeq7.5\times10^{-16}$ and $2.4\times10^{-16}$ for signal modulation at 2 and 20\,mHz, respectively, below current ground-based torsion-balance limits.
For a baryon-coupled Yukawa interaction sourced by the Earth, the corresponding projected sensitivity to the coupling strength is $\alpha\simeq3.1\times10^{-13}$ and $9.8\times10^{-14}$ for ranges $\lambda\gtrsim10^{7}$\,m. We identify the modulation scheme, the material contrast, and the source geometry as the main factors that set the projected reach, and the control of systematic couplings as the main experimental requirement for attaining it. These results indicate that differential torsion sensors are a promising platform for future laboratory WEP tests.

\end{abstract}

\maketitle

\section{Introduction}
\label{sec:intro}

The weak equivalence principle (WEP), which states that all freely falling test bodies experience the same gravitational acceleration independent of their internal composition, is a cornerstone of general relativity. Any violation of the WEP would imply the existence of interactions beyond pure metric gravity and would provide evidence for new physics beyond both general relativity and the Standard Model~\cite{Adelberger2009, Wagner_2012, Will2014}. A broad class of extensions of fundamental physics, including light scalar and vector fields, dilaton-like interactions, and theories containing additional gauge symmetries, naturally predict composition-dependent forces that manifest as deviations from universal free fall~\cite{Damour2012,Uzan2025,Graham2016}. Many such theories introduce new mediators that couple to quantities such as baryon number or baryon-minus-lepton number, generating differential accelerations between materials with distinct compositions~\cite{Fischbach1992,Adelberger2003,Wagner_2012}. Consequently, precision tests of the WEP remain among the most powerful experimental probes of new long-range interactions.

Over the past several decades, increasingly sensitive experiments have pushed tests of the WEP to unprecedented precision. Ground-based rotating torsion balances developed by the E\"ot-Wash collaboration have constrained composition-dependent differential acceleration
to a level corresponding to $\eta\sim10^{-13}$~\cite{Schlamminger2008,Wagner_2012}, while lunar laser ranging experiments have achieved comparable precision~\cite{williams2012lunar,muller2012testing}. Torsion balances have also been used to test the WEP for superconducting test bodies~\cite{ross2025test}.
More recently, the MICROSCOPE satellite mission~\cite{Touboul2017,Touboul2022} improved the limit to approximately $\eta \sim 10^{-15}$ using drag-free differential accelerometers in orbit. In parallel, atom-interferometric approaches are being actively developed as a complementary route toward future high-precision tests of the WEP~\cite{aguilera2014ste,struckmann2024platform}. Although these experiments have constrained large regions of parameter space~\cite{Adelberger2003,Fischbach1998}, substantial regions remain to be explored, motivating next-generation experiments~\cite{Berge2018}.

Recent developments in precision torsion sensing, particularly differential torsion configurations employing optical cavity readout, have opened a promising path toward substantially improved force sensitivity~\cite{Chua2023}.
Originally developed for low-frequency Newtonian-noise detection, the differential torsion sensor combines long mechanical coherence times and high-sensitivity optical readout~\cite{Chua2023}. Compared with conventional single-pendulum torsion balances, the differential configuration can suppress disturbances that couple similarly to both oscillators while retaining sensitivity to differential forces. The same sensor architecture has subsequently been proposed for the direct detection of ultralight vector dark matter~\cite{Sun2025}. Related torsion-pendulum approaches have also been explored for ultralight dark-matter searches; in particular, a double-torsion-pendulum scheme employing cross-correlated readout has been proposed to suppress suspension thermal noise and improve sensitivity at millihertz frequencies~\cite{okuma2025cross}.

These developments motivate the application of differential torsion sensing to tests of the WEP and searches for composition-dependent interactions. The use of test masses with distinct compositions, such as Be--Ti or Be--Al~\cite{Adelberger2009,Wagner_2012}, enables sensitivity to composition-dependent forces through the differential accelerations they induce. In the dark-matter case of Ref.~\cite{Sun2025}, the signal oscillates at the Compton frequency of the field. A WEP test with the Earth as the source differs in that the signal is static in the laboratory, so it must be modulated externally, and its amplitude depends on the finite-range Earth source and on the horizontal projection at the site. The present work addresses these aspects for the same instrument and noise model. In this work, we adopt Be--Ti as the reference material pair because of its substantial composition contrast and its established use in previous WEP experiments~\cite{Schlamminger2008}. We evaluate the projected sensitivity of a differential torsion sensor to such interactions using representative instrument parameters and realistic noise models. We derive projected sensitivities to differential acceleration and the E\"otv\"os parameter, and translate these into projected sensitivities to Yukawa-type fifth-force interactions. Our results indicate that such sensors could provide competitive laboratory sensitivity for interaction ranges $\lambda\gtrsim10^{7}$\,m.

The remainder of this paper is organized as follows.
Section~\ref{sec:signal_model} introduces the E\"otv\"os parameter, Yukawa parameterization, differential-torque formulation, and torsional response of the sensor.
Section~\ref{sec:sensitivity} develops the sensitivity formalism under both static and modulated measurement scenarios.
Section~\ref{sec:sensor} describes the representative differential torsion sensor and its noise performance.
Section~\ref{sec:results} presents the resulting sensitivities to differential acceleration, the E\"otv\"os parameter, and Yukawa-type interactions.
Section~\ref{sec:discussion} discusses the principal factors governing experimental reach, including sensitivity scaling, material dependence, and potential systematic limitations.
Finally, Section~\ref{sec:conclusion} summarizes the main results and outlines future applications of differential torsion sensors to tests of fundamental physics.

\section{Signal Model for WEP Tests}\label{sec:signal_model}

WEP violation is conventionally quantified by the E\"otv\"os parameter~\cite{Will2014,Wagner_2012}

\begin{equation}
\eta \equiv 2\frac{|\mathbf{a}_A-\mathbf{a}_B|}{|\mathbf{a}_A+\mathbf{a}_B|},
\end{equation}
where $\mathbf{a}_A$ and $\mathbf{a}_B$ are the free-fall accelerations of
two test bodies $A$ and $B$, of different composition
toward the same gravitational source. Defining the differential and common acceleration magnitudes as
$\Delta a \equiv |\mathbf{a}_A-\mathbf{a}_B|$ and
$a \equiv \frac{1}{2}|\mathbf{a}_A+\mathbf{a}_B|$ respectively,
the E\"otv\"os parameter can be written as
\begin{equation}
    \eta = \frac{\Delta a}{a}.
    \label{eq:eta_approx}
\end{equation}

To connect this phenomenological parameter to an underlying interaction model, we use the standard Yukawa parameterization of a composition-dependent fifth force~\cite{Adelberger2009,Wagner_2012,Fischbach1998},

\begin{equation}
V_{Y}(r)= -\,G\,m_i m_j \,\alpha\,\left(\frac{q_i}{\mu_i}\right)\!\left(\frac{q_j}{\mu_j}\right)\,\frac{e^{-r/\lambda}}{r},
\label{eq:yukawa-potential}
\end{equation}
where $V_Y(r)$ is the Yukawa interaction potential between bodies $i$ and $j$, $G$ is Newton's gravitational constant, $m_i$ and $m_j$ are their masses, and $r$ is their separation. The dimensionless parameter $\alpha$ characterizes the interaction strength relative to gravity, $\lambda$ is the
interaction range, and $q/\mu$ denotes the effective new
charge per atomic mass unit, where $q$ is the charge associated with the hypothetical interaction and $\mu$ is the atomic mass in amu. Depending on the underlying theory, the charge may be proportional to baryon number $B$, proton number $Z$, neutron number $N$, or the combination $B-L$ with $L$ the lepton number. The derivation is summarized in Appendix~\ref{app:yukawa}.

For a differential torsion sensor with test bodies of different composition, the component of the Yukawa-induced differential acceleration along the sensor's sensitive axis can be written as
\begin{equation}
\Delta a_Y(\lambda) = G\,\alpha(\lambda)\,|\Delta(q/\mu)|
\,(q/\mu)_{\rm src}\,S(\lambda)\,\mathcal{P},
\label{eq:yukawa_acceleration}
\end{equation}

where $\Delta(q/\mu) = (q/\mu)_A-(q/\mu)_B$ is the material contrast, $(q/\mu)_{\rm src}$ is the corresponding charge-to-mass ratio of the
source, $S(\lambda)$ is a source factor that encapsulates the geometry, density distribution, and finite-range effects associated with the attracting mass, and $\mathcal{P}$ is a geometric factor accounting for the projection of the differential acceleration onto the sensor's sensitive axis. The derivation of $\mathcal{P}$ is provided in Appendix~\ref{app:yukawa}.

\begin{figure}[h]
  \centering
\includegraphics[width=0.8\linewidth]{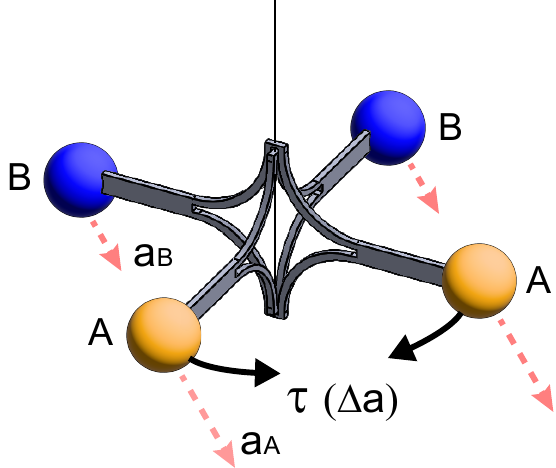}
  \caption{Schematic illustration of the differential torsion sensor used for WEP tests. Two torsion pendulums, suspended in a cross formation from a common suspension point, each carry equal-mass test masses of compositions $A$ and $B$ at the two ends of the beam. A composition-dependent (Yukawa-type) interaction produces differential accelerations $\mathbf{a}_A$ and $\mathbf{a}_B$ on the two test-mass compositions, resulting in a differential torque $\tau(\Delta \mathbf{a})$ about the suspension axis and a measurable torsional response. In this work, compositions $A$ and $B$ correspond to titanium (Ti) and beryllium (Be), the reference material pair adopted for the sensitivity estimates.}
  \label{fig:overview}
\end{figure}

The overall signal generation mechanism is illustrated schematically in Fig.~\ref{fig:overview}. The differential acceleration signal is converted into a measurable
torque through the torsional response of the sensor. In the frequency
domain, the torsional angle $\theta(f)$ is related to the applied
torque $\tau(f)$ by
\begin{equation}
\theta(f)=H(f)\tau(f),
\end{equation}
where the torsional transfer function with structural damping~\cite{Saulson1990} is
\begin{equation}
H(f)
=
\frac{1}{\kappa-(2\pi f)^2I+i\kappa\phi},
\label{eq:transfer_function}
\end{equation}
where $f$ is the Fourier frequency, $I$ is the moment of inertia about the torsional axis, $\kappa$ is the torsional spring constant, $\phi=Q^{-1}$ is the mechanical loss angle, $Q$ is the mechanical quality factor, and $i$ is the imaginary unit.

The Yukawa interaction produces a differential acceleration and hence a signal torque, which is measured through the angular displacement of the torsion sensor.

\section{SENSITIVITY FORMALISM}
\label{sec:sensitivity}
For a differential force acting perpendicular to the lever arm, the resulting torque is
\begin{equation}
\tau_{\rm sig}=Mr\Delta a_{\rm sig},
\label{eq:signal_torque}
\end{equation}
\noindent
where \(M\) is the total mass of each test-mass material (\(M=10\)\,kg, i.e., two 5-kg test masses), following the convention of Ref.~\cite{Sun2025}, and \(r\) is the lever arm length.

Let \(S_\tau(f)\) denote the one-sided torque-noise power spectral density, including suspension thermal noise, optical readout noise, and other technical noise contributions. The corresponding differential-acceleration noise amplitude spectral density is
\begin{equation}
S_a^{1/2}(f)
=
\frac{S_\tau^{1/2}(f)}{Mr}.
\label{eq:acceleration_noise}
\end{equation}

For a signal at frequency \(f\), the signal-to-noise ratio (SNR) after an effective integration time \(T_{\rm eff}\) is
\begin{equation}
{\rm SNR}
=
\frac{\Delta a_{\rm sig}}
{S_a^{1/2}(f)}
\sqrt{T_{\rm eff}}.
\label{eq:snr}
\end{equation}
By setting the SNR to unity, we obtain the differential-acceleration sensitivity,
\begin{equation}
\Delta a_{\rm sens}(f)
=
\frac{S_a^{1/2}(f)}
{\sqrt{T_{\rm eff}}}.
\label{eq:acceleration_sensitivity}
\end{equation}
In practice, a confident detection would generally require a higher SNR threshold, depending on the noise statistics and false-alarm probability.

Using Eq.~(\ref{eq:eta_approx}), the corresponding sensitivity to the E\"otv\"os parameter is
\begin{equation}
\eta_{\rm sens}(f)
=
\frac{\Delta a_{\rm sens}(f)}{a_{\parallel}}
=
\frac{S_a^{1/2}(f)}
{g\,\sqrt{T_{\rm eff}}\,\mathcal{P}}.
\label{eq:eta_sens}
\end{equation}
Here, the common acceleration projected onto the sensor plane is
$a_{\parallel}\simeq g\mathcal{P}$, where $g$ is the local gravitational acceleration
and $\mathcal{P}$ is the projection factor from Sec.~\ref{sec:signal_model}.

For a Yukawa-type composition-dependent interaction, Eq.~(\ref{eq:yukawa_acceleration}) relates the differential acceleration to the coupling strength \(\alpha\). The corresponding sensitivity to \(\alpha\) is
\begin{equation}
\alpha_{\rm sens}(\lambda,f)
=
\frac{\Delta a_{\rm sens}(f)}
{G|\Delta(q/\mu)|\,(q/\mu)_{\rm src}\,S(\lambda)\,\mathcal{P}}.
\label{eq:alpha_sens}
\end{equation}
Thus, the differential-acceleration sensitivity can be expressed either as a sensitivity to the E\"otv\"os parameter or, for a specified interaction range and source geometry, as a sensitivity to the Yukawa coupling strength.

Two measurement strategies are considered: static and modulated measurements. In a static measurement, a possible WEP-violating signal appears as a DC torque.
Assuming stationary, noise-limited operation over the full observing time, the effective integration time equals the total observation time $T_{\rm obs}$, i.e., $T_{\rm eff}=T_{\rm obs}$. Because $S_a^{1/2}(f)$ rises toward low frequencies (Fig.~\ref{fig:acc_spectrum}), a purely static measurement is limited by low-frequency noise and drifts; the sensitivities below are therefore evaluated for signals at nonzero modulation frequencies.

In practice, low-frequency drifts and environmental systematics often limit coherent integration over long continuous observation times. For this reason, many torsion-balance experiments employ signal modulation, shifting the signal to a cleaner frequency band where lock-in techniques can be applied. For a modulated measurement, the modulation phase is externally controlled and therefore the signal, in principle, is coherent. The coherent integration may instead be limited by the stability of the measurement, including the modulation phase and detector response.
The effective coherence time is therefore \(T_{\rm coh}=\min(T_{\rm meas},T_{\rm obs})\),
where \(T_{\rm meas}\) is the timescale over which the modulation
phase and detector response remain sufficiently stable for coherent
integration. If \(T_{\rm meas}\geq T_{\rm obs}\), the signal can be
integrated coherently over the full observing time and
\(T_{\rm eff}=T_{\rm obs}\). If \(T_{\rm meas}<T_{\rm obs}\), coherent
segments of duration \(T_{\rm coh}\) may be combined incoherently,
giving \(T_{\rm eff}\simeq\sqrt{T_{\rm obs}T_{\rm coh}}\). This applies when the relative phase between segments cannot be recovered; with an external phase reference, the segments can still be combined coherently.
For a source fixed in the laboratory frame, such as the Earth, the signal must be moved to $f_{\rm mod}$ by rotating the sensor relative to the source; the implications are discussed in Sec.~\ref{sec:systematics}.

\section{Representative Differential Torsion Sensor}
\label{sec:sensor}

To evaluate the achievable sensitivity, we consider the representative
differential torsion sensor parameters listed in
Table~\ref{tab:sensor_parameters}. Some of these parameters are adopted
from the instrument described in Ref.~\cite{Chua2023}, while others
are chosen as representative values for the sensitivity analysis. We assume a suspension quality factor of $Q=10^{5}$, representative of fused-silica torsion fibres~\cite{Cavalleri2009.CQG}.
\begin{table}[h]
\caption{Representative differential torsion sensor parameters used
for the sensitivity estimates.}
\label{tab:sensor_parameters}
\centering
\begin{tabular}{lc}
\hline\hline
Parameter & Value \\
\hline
Test mass (each end) & 5 kg \\
Total mass per material $M$ & 10 kg \\
Effective lever arm length $r$ & 0.24 m \\
Moment of inertia $I$ & 0.75 kg\,m$^2$ \\
Torsional resonant frequency & 0.026 Hz \\
Quality factor $Q$ & $10^5$ \\
Cavity finesse & 150 \\
Carrier power $P_c$ & 10 mW \\
Laser wavelength & 1064 nm \\
Shot-noise level & $2\times10^{-17}$ rad Hz$^{-1/2}$ \\
\hline\hline
\end{tabular}
\end{table}

The instrument comprises two torsion pendulums arranged in a cross
configuration with co-located centers of mass and a common suspension
point. Each beam carries equal-mass test masses of the two materials (here Be and Ti) at its two ends, as in Ref.~\cite{Sun2025}.
This mechanical design provides rejection of disturbances that couple commonly to the two pendulums, and their differential rotation is measured using an optical cavity readout. The instrument is operated on a multi-stage
seismic-isolation platform under vacuum. We assume residual
suspension-point motion of
$10^{-11}\,\mathrm{m\,Hz^{-1/2}}$ and
$10^{-11}\,\mathrm{rad\,Hz^{-1/2}}$ at 0.1 Hz, consistent with
Ref.~\cite{Chua2023}. Under these assumptions, the sensitivity is
dominated by suspension thermal noise across the most sensitive
frequency band.

\begin{figure}[t]
  \centering
\includegraphics[width=0.49\textwidth]{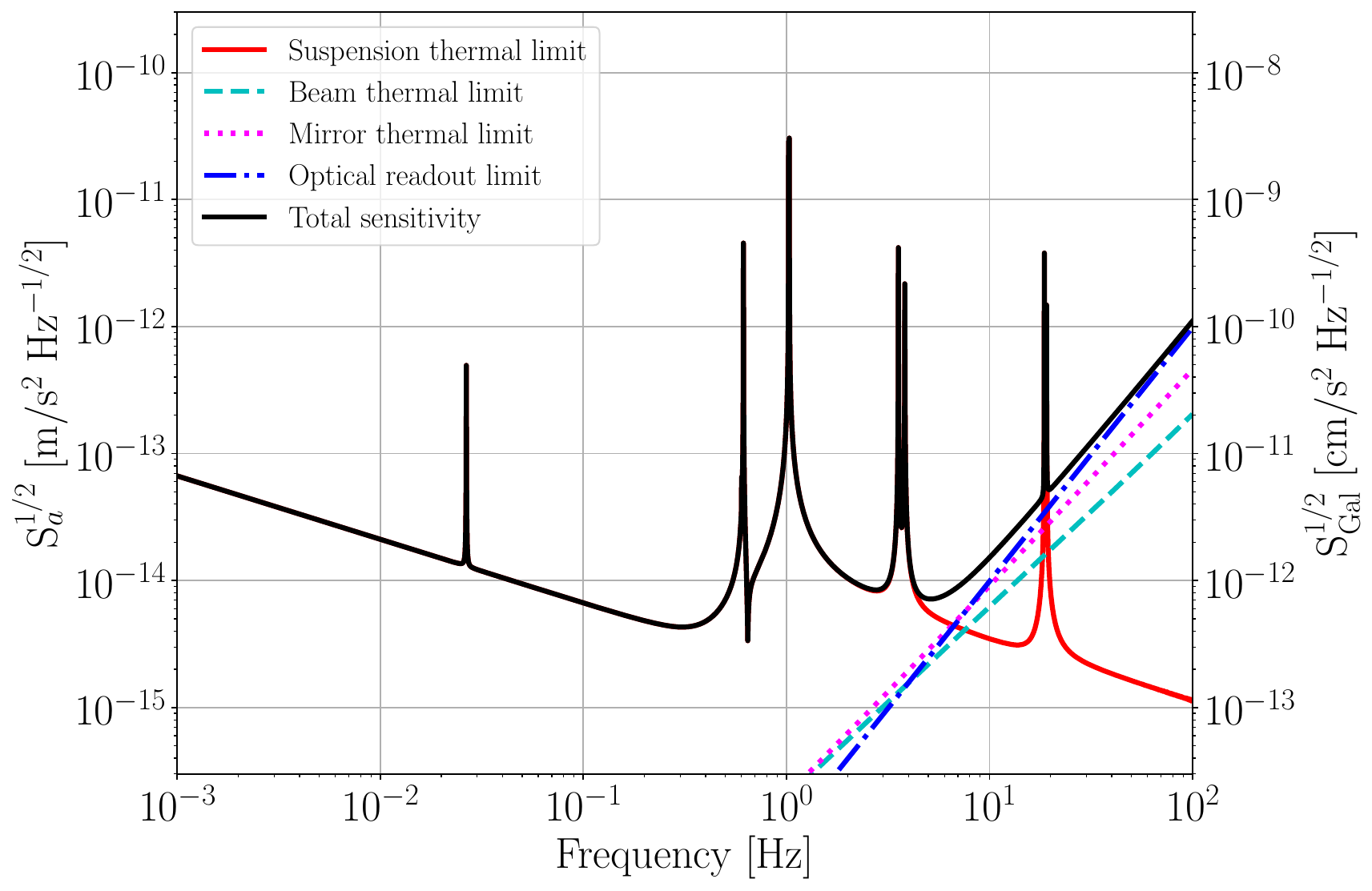}
  \caption{Differential-acceleration sensitivity of the representative differential torsion sensor. The total sensitivity includes suspension thermal, beam thermal, mirror thermal and optical readout limits. Resonance peaks are from thermal motion in the sensor's nominal suspension translational and rotational degrees-of-freedom. The suspension thermal noise limit dominates below $\sim 5\,{\rm Hz}$, while optical readout noise limits the high-frequency floor.}
  \label{fig:acc_spectrum}
\end{figure}

The differential angular motion is measured using Fabry--P\'erot
cavity readout. For the parameters listed in
Table~\ref{tab:sensor_parameters}, the displacement-equivalent shot
noise is estimated to be
$4\times10^{-18}~{\rm m\,Hz^{-1/2}}$, corresponding to an angular
sensitivity of approximately
$2\times10^{-17}~{\rm rad\,Hz^{-1/2}}$.
The resulting differential-acceleration noise budget is shown in
Fig.~\ref{fig:acc_spectrum}. The total noise model includes suspension
thermal noise, beam thermal noise, mirror thermal noise, and optical readout noise. Expressions for the suspension and beam thermal noise and for the shot-noise contribution to the readout are given in Appendix~\ref{appx:noise_sources}; the mirror thermal noise model follows Ref.~\cite{Sun2025}. The suspension thermal noise of both pendulums is included. It is taken from the multi-degree-of-freedom model of the prototype suspension~\cite{Chua2023,Forsyth-thesis}, as in Ref.~\cite{Sun2025}, and rescaled from the model value $Q=10^{6}$ to $Q=10^{5}$ assuming structural damping ($S_\tau\propto Q^{-1}$). Below 0.01\,Hz, outside the range of the model, it is extended using Eq.~(\ref{eq:thermal}) normalized to the model at 0.01\,Hz; the 2\,mHz baseline adopted below lies in this extended range. The projections below therefore adopt the nominal response and noise normalization of this model and are estimates for a representative configuration; an independent calibration for a rotating WEP configuration is beyond the scope of this work.
For convenience, the right vertical axis is also expressed in Gal
($1~\mathrm{Gal}=1~\mathrm{cm\,s^{-2}}$), a unit commonly used in
gravimetry.

The resulting acceleration noise spectrum $S_a^{1/2}(f)$ provides
the input to the sensitivity formalism developed in
Sec.~\ref{sec:sensitivity}, from which the differential-acceleration,
E\"otv\"os-parameter, and Yukawa-interaction sensitivities are
evaluated in the following section.

\section{WEP and Fifth-Force Sensitivity}
\label{sec:results}

\begin{figure}[t]
  \centering
  \includegraphics[width=0.45\textwidth]{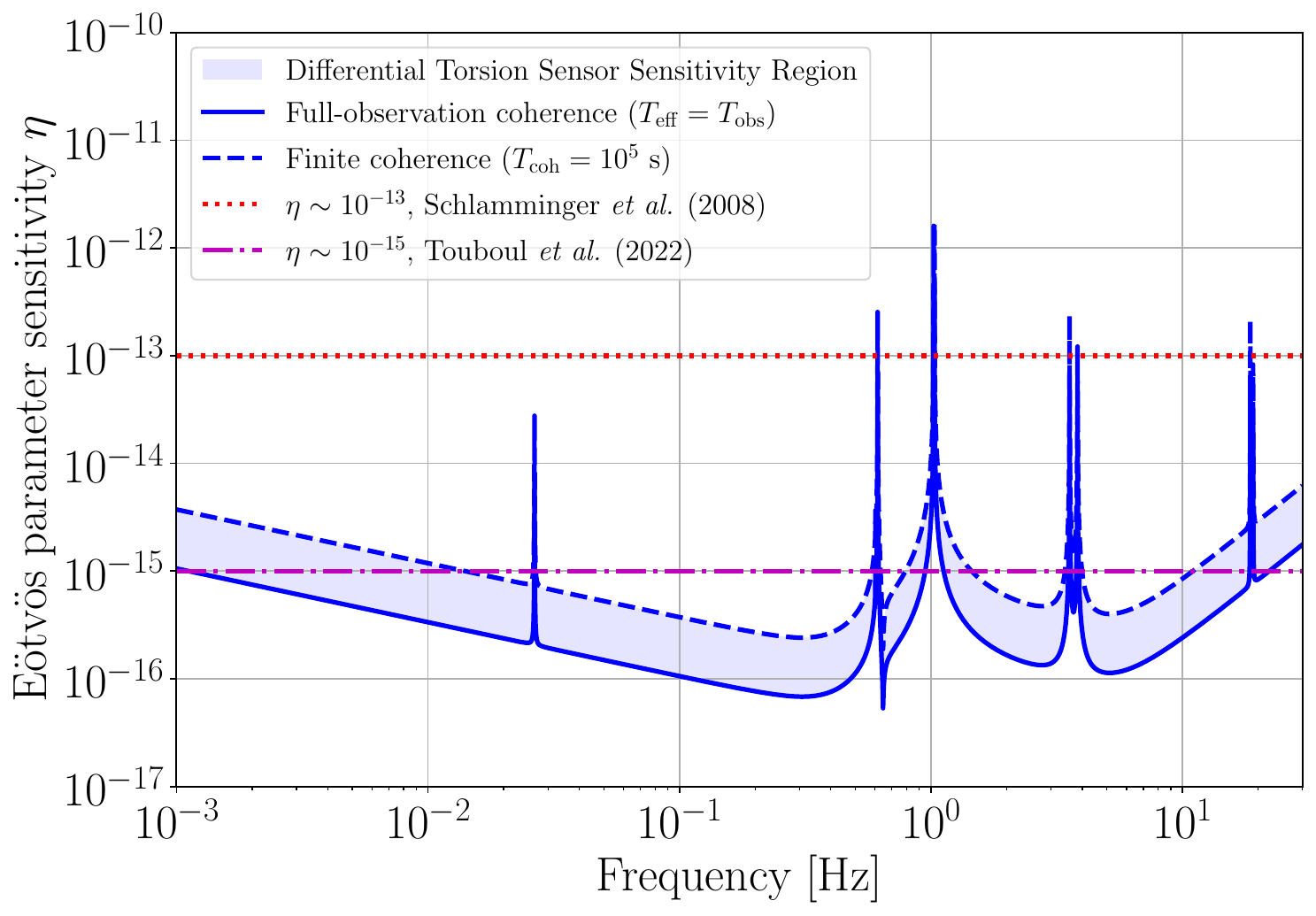}
\caption{Projected sensitivity to the E\"otv\"os parameter $\eta(f)$
as a function of frequency, derived from the differential-acceleration
noise spectrum using Eq.~(\ref{eq:eta_sens}). The solid and dashed curves correspond to the two integration assumptions defined in Sec.~\ref{sec:sensitivity}, $T_{\rm eff}=T_{\rm obs}$ and $T_{\rm eff}\simeq\sqrt{T_{\rm obs}T_{\rm coh}}$, with \(T_{\mathrm{obs}}=180\) days and \(T_{\mathrm{coh}}=10^{5}\,\mathrm{s}\). The shaded region spans the range between them. The projection factor is
$\mathcal{P}=1.63\times10^{-3}$ for Canberra, Australia (Appendix~\ref{app:yukawa}). Horizontal lines indicate the approximate levels ($\eta\sim10^{-13}$ and $\sim10^{-15}$) of existing limits from Refs.~\cite{Schlamminger2008} and \cite{Touboul2022}. The horizontal axis is the signal frequency, i.e., the modulation frequency $f_{\rm mod}$ for an Earth-fixed source.}
  \label{fig:eta_spectrum}
\end{figure}

The sensitivity to the E\"otv\"os parameter $\eta(f)$, obtained from
Eq.~(\ref{eq:eta_sens}), is shown in Fig.~\ref{fig:eta_spectrum}.
The solid curve assumes coherent integration over the full observing time,
$T_{\rm eff}=T_{\rm obs}$, with $T_{\rm obs}=180$ days. The dashed curve illustrates a finite-coherence
case with $T_{\rm coh}=10^5\,{\rm s}$ and
$T_{\rm eff}\simeq\sqrt{T_{\rm obs}T_{\rm coh}}$.
The shaded region spans these two integration assumptions.
For a modulated measurement at modulation frequency $f_{\rm mod}$, the signal is shifted to $f_{\rm mod}$ and the
sensitivity is determined by $S_a^{1/2}(f_{\rm mod})$; if coherence is
maintained over the full observing time, the modulated measurement also
reaches the solid-curve value.
The value $T_{\rm coh}=10^5\,{\rm s}$ is illustrative rather than an
experimental limit.
The projected sensitivity improves from $\eta\approx10^{-15}$ at 1\,mHz to below $10^{-16}$ between about 0.1 and 0.6\,Hz; the approximate levels of existing limits~\cite{Schlamminger2008,Touboul2022} are shown for reference. At the modulation frequencies adopted below (Sec.~\ref{sec:systematics}), the full-coherence values are $\eta\simeq7.5\times10^{-16}$ at 2\,mHz and $2.4\times10^{-16}$ at 20\,mHz.

\begin{figure}[t]
  \centering
  \includegraphics[width=0.48\textwidth]{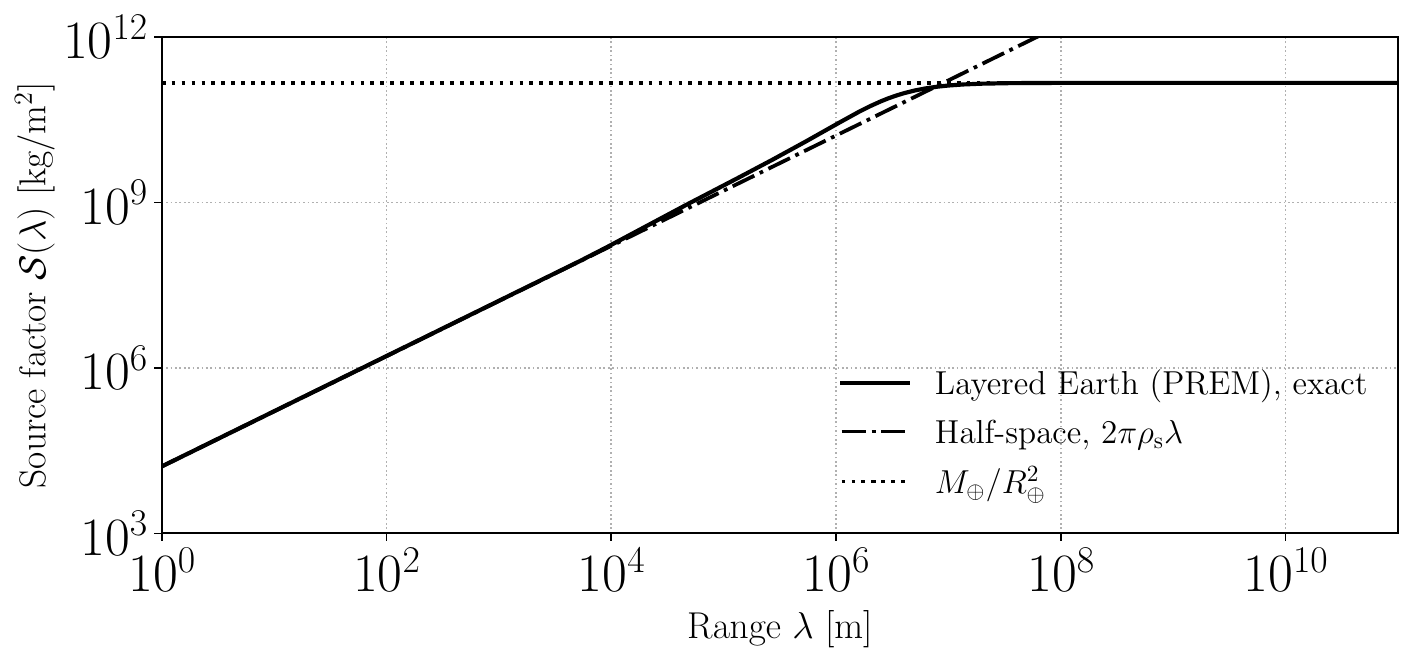}
  \caption{Yukawa source factor $S(\lambda)$ at the Earth's surface for the PREM density profile, Eq.~(\ref{eq:S_exact}) (solid), with its short-range half-space limit $2\pi\rho_{\rm s}\lambda$ ($\rho_{\rm s}=2.6\,\mathrm{g\,cm^{-3}}$) and long-range limit $M_\oplus/R_\oplus^{2}$.}

  \label{fig:source_models}
\end{figure}

To express the projected E\"otv\"os-parameter sensitivity in terms of the Yukawa coupling strength, the source factor $S(\lambda)$ must be specified. For a spherically symmetric Earth, $S(\lambda)$ follows exactly from the Yukawa analogue of the shell theorem (Appendix~\ref{app:source})~\cite{Fischbach1998,Fayet2025,Fayet2026},
\begin{equation}
S(\lambda)=\frac{(1+R_\oplus/\lambda)e^{-R_\oplus/\lambda}}{R_\oplus^{2}}
4\pi\lambda\!\int_0^{R_\oplus}\!\rho(r)\,r\sinh(r/\lambda)\,dr,
\label{eq:S_exact}
\end{equation}
where $\rho(r)$ is the radial density profile. This expression has no adjustable parameters and interpolates between two familiar limits: for $\lambda\ll R_\oplus$ only material within $\sim\lambda$ of the surface contributes and $S\to2\pi\rho_{\rm s}\lambda$, the half-space result for a near-surface density $\rho_{\rm s}$~\cite{Adelberger2009,Fischbach1998}; for $\lambda\gg R_\oplus$ it saturates at the unscreened value $S_\infty=M_\oplus/R_\oplus^{2}$. Figure~\ref{fig:source_models} shows $S(\lambda)$ for the preliminary reference Earth model (PREM)~\cite{Dziewonski1981}. At intermediate ranges the denser deep layers raise $S(\lambda)$ above the half-space extrapolation by up to a factor of $\simeq1.7$.

We adopt the projection factor of Eq.~(\ref{eq:proj_factor}) as a long-range approximation, appropriate for interaction ranges comparable to or larger than the Earth's radius. At shorter ranges, a torsion balance on flat, level ground has essentially no sensitivity to an Earth-sourced Yukawa force, and the horizontal field is set by the local topography and laboratory environment, which requires a site-specific source model~\cite{Adelberger1990,Wagner_2012,Schlamminger2008}. We therefore quote Earth-source projections only for $\lambda\gtrsim10^{7}$\,m.

\begin{figure}[t]
  \centering
  \includegraphics[width=0.48\textwidth]{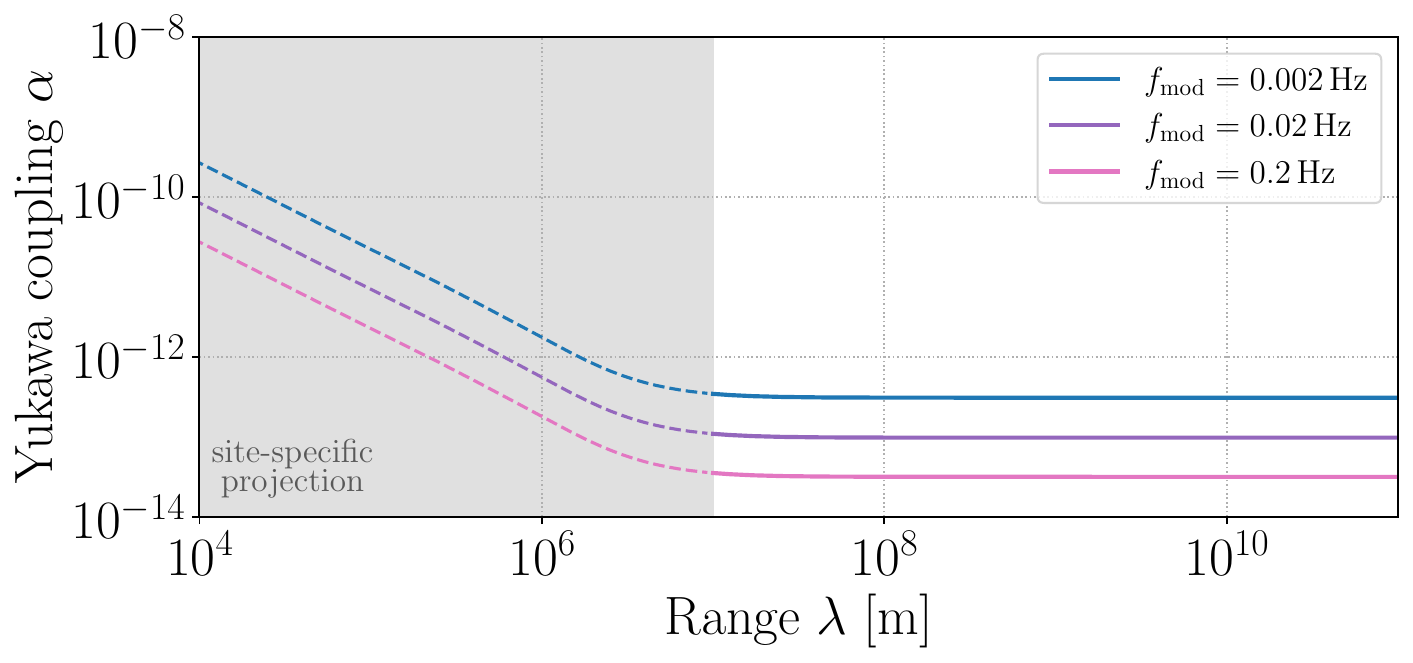}
\caption{Projected sensitivity to the Yukawa coupling $\alpha(\lambda)$ for a baryon-coupled interaction with the Earth as the source (Be--Ti, $|\Delta(B/\mu)|=2.42\times10^{-3}$), for the three modulation frequencies shown in the legend. Each curve uses Eq.~(\ref{eq:alpha_sens_fmod}) with the acceleration noise evaluated at $f_{\rm mod}$, $T_{\rm eff}=T_{\rm obs}=180$\,days, $S(\lambda)$ of Fig.~\ref{fig:source_models}, and $\mathcal{P}=1.63\times10^{-3}$. In the shaded region ($\lambda<10^{7}$\,m) the horizontal Yukawa field depends on the local topography and laboratory environment; the dashed continuation is shown for reference only and is not a projected sensitivity.}

  \label{fig:alpha_lambda}
\end{figure}

Combining the acceleration sensitivity with the source factor defined above,
the corresponding sensitivity to a Yukawa-type interaction can be written as

\begin{equation}
\alpha(\lambda,f)=
\frac{S_a^{1/2}(f)}{\sqrt{T_{\rm eff}}}
\,
\frac{1}{
G\,|\Delta(q/\mu)|\,(q/\mu)_{\rm src}\,S(\lambda)\,\mathcal{P}}.
\label{eq:alpha_sens_fmod}
\end{equation}
For the Be--Ti test-mass and baryon coupling (\(q=B\)), the charge-to-mass contrast is
$|\Delta(B/\mu)|=2.42\times10^{-3}$ (Table~\ref{tab:baryon_contrast}).
For the baryon coupling considered here, the Earth's source
charge-to-mass ratio is approximately unity,
$(B/\mu)_{\rm src}\simeq1$, which is adopted in the following
sensitivity estimates. Because $GS_\infty\simeq g$, Eqs.~(\ref{eq:eta_sens}) and (\ref{eq:alpha_sens_fmod}) give $\alpha\simeq\eta_{\rm sens}/[|\Delta(q/\mu)|(q/\mu)_{\rm src}]$ at long range, which connects Figs.~\ref{fig:eta_spectrum} and~\ref{fig:alpha_lambda} directly.

For the modulated measurement, we evaluate
Eq.~(\ref{eq:alpha_sens_fmod}) at three representative modulation frequencies (2\,mHz as the baseline turntable rate, 20\,mHz as a higher-frequency benchmark, and 0.2\,Hz as an optimistic case; see Sec.~\ref{sec:systematics}), assuming coherent
integration over the full observing time, $T_{\rm eff}=T_{\rm obs}$.
The resulting sensitivities are shown in Fig.~\ref{fig:alpha_lambda}.
At long interaction ranges ($\lambda\gtrsim10^7\,{\rm m}$), the projected
sensitivities approach $\alpha\simeq3.1\times10^{-13}$,
$9.8\times10^{-14}$, and $3.2\times10^{-14}$ for
$f_{\rm mod}=2$\,mHz, 20\,mHz, and 0.2\,Hz, respectively; with the long-range projection factor adopted here, the values at $\lambda=10^{7}$\,m are about $12\%$ higher.
For comparison, the published long-range baryon-coupling constraint of Ref.~\cite{Schlamminger2008} is $\alpha\sim10^{-10}$; it includes systematic uncertainties, whereas the values above are noise-limited (SNR\,$=1$) projections that depend on the modulation scheme discussed in Sec.~\ref{sec:systematics}.

\section{Discussion}
\label{sec:discussion}

\subsection{Sensitivity Scaling and Design Levers}

To quantify the relative impact of different experimental parameters,
we define the logarithmic sensitivity index

\begin{equation}
\chi_x \equiv
\frac{\partial \ln \eta}
{\partial \ln x},
\end{equation}
\noindent
which measures the fractional change in sensitivity resulting from
a fractional change in parameter \(x\).

For coherent integration over the full observing time,
\(T_{\rm eff}=T_{\rm obs}\), Eq.~(\ref{eq:eta_sens}) gives
$\chi_{S_a^{1/2}}=1$ and $\chi_{T_{\rm obs}}=-1/2$.
In the finite-coherence case,
\(T_{\rm eff}\simeq\sqrt{T_{\rm obs}T_{\rm coh}}\), $\chi_{S_a^{1/2}}$ is unchanged and $\chi_{T_{\rm obs}}=\chi_{T_{\rm coh}}=-1/4$.
These indices provide a simple measure of experimental leverage.
For example, reducing the differential acceleration-noise floor
by a factor of ten improves the E\"otv\"os sensitivity by the same
factor, whereas achieving an equivalent improvement through
integration alone requires a hundred-fold increase in observing
time under full coherence. In the finite-coherence regime, the
same improvement would require a ten-thousand-fold increase in
either $T_{\rm obs}$ or $T_{\rm coh}$.
For the sensitivity estimates presented here, we take
$T_{\rm obs}=180$ days; $T_{\rm coh}=10^5\,{\rm s}$ in Fig.~\ref{fig:eta_spectrum}
is used only to illustrate a possible finite-coherence case rather
than as an assumed experimental limit.

The test-mass scale is a further design lever. For fixed suspension stiffness, loss angle, and temperature, the thermal torque noise $S_\tau$ is independent of the test mass, while the signal torque scales as $Mr$ [Eq.~(\ref{eq:signal_torque})], so that $\chi_M=\chi_r=-1$ in the thermal-noise-limited band. In practice the suspension load and the handling of large beryllium masses (5\,kg of Be occupies $\simeq2.7$\,L) limit how far this can be pushed.

\subsection{Material Dependence}

The sensitivity to Yukawa-type interactions depends not only on the
instrumental noise floor but also on the charge-to-mass contrast between
the test masses and the corresponding charge-to-mass ratio of the
source. From Eq.~(\ref{eq:alpha_sens}),

\begin{equation}
\alpha_{\rm sens}
\propto
\left[
|\Delta(q/\mu)|\,(q/\mu)_{\rm src}
\right]^{-1}.
\end{equation}
Material pairs with larger charge-to-mass contrast therefore provide
stronger sensitivity for a given source composition.

Tables~\ref{tab:baryon_contrast} and~\ref{tab:neutron_contrast}
summarize the absolute contrast factor
$|\Delta(q/\mu)|$ for several representative material combinations
under baryon-number and neutron coupling, respectively. The elemental
charge-to-mass ratios are calculated using representative natural
isotopic abundances~\cite{meija2016isotopic} and standard atomic
weights~\cite{prohaska2022standard}. Specifically, for an element
with atomic number $Z$,
\begin{equation}
    \overline{B}=\sum_i w_i A_i,
    \qquad
    \overline{N}=\overline{B}-Z,
\end{equation}
where $w_i$ and $A_i$ are the natural abundance and mass number of
isotope $i$, respectively. The corresponding elemental ratios are
$\overline{B}/\overline{\mu}$ and
$\overline{N}/\overline{\mu}$, where $\overline{\mu}$ is the standard
atomic weight.

The Be--Ti pair is adopted as the reference configuration because it
provides a substantial composition contrast and has been widely used
in previous equivalence-principle experiments~\cite{Schlamminger2008}. For baryon
coupling, $(B/\mu)_{\rm src}\simeq 1$ is assumed for the source. This
approximation is appropriate for the bulk source term, but the small
material-dependent deviations of $B/\mu$ from unity are retained when
calculating the test-mass contrast $\Delta(B/\mu)$. For neutron
coupling, we use an effective bulk-Earth value
$(N/\mu)_{\rm src}\simeq0.51$, estimated from the elemental
composition of the Earth~\cite{mcdonough1995composition}. All relative $\alpha$ sensitivities in Tables~\ref{tab:baryon_contrast} and~\ref{tab:neutron_contrast} are normalized to the Be--Ti pair under baryon coupling.

\begin{table}[h]
\caption{Material charge-to-mass contrast $|\Delta(B/\mu)|$ and
relative projected $\alpha$ sensitivity for representative material
pairs under baryon coupling. The relative sensitivity is normalized
to the Be--Ti pair.}
\begin{ruledtabular}
\begin{tabular}{lccc}
Materials &
$|\Delta(B/\mu)|$ &
Rel.\ $\alpha$ limit &
$\log_{10}(\mathrm{Rel.}\ \alpha)$ \\
\hline
Be--Ti & $2.42\times10^{-3}$ & 1.00 & $0.00$ \\
Be--Al & $2.04\times10^{-3}$ & 1.19 & $0.08$ \\
Be--W  & $1.63\times10^{-3}$ & 1.49 & $0.17$ \\
Al--Ti & $3.87\times10^{-4}$ & 6.25 & $0.80$ \\
Al--W  & $4.08\times10^{-4}$ & 5.93 & $0.77$ \\
Ti--W  & $7.96\times10^{-4}$ & 3.04 & $0.48$ \\
\end{tabular}
\end{ruledtabular}
\label{tab:baryon_contrast}
\end{table}

\begin{table}[h]
\caption{Material charge-to-mass contrast $|\Delta(N/\mu)|$ and relative $\alpha$
sensitivity for various material pairs under neutron coupling.
The effective bulk-Earth neutron charge-to-mass ratio is taken as
$(N/\mu)_{\rm src}\simeq0.51$, estimated from the bulk-Earth
composition of Ref.~\cite{mcdonough1995composition}. Sensitivity is normalized
to the Be--Ti baseline under baryon coupling.}
\begin{ruledtabular}
\begin{tabular}{lccc}
Materials &
$|\Delta(N/\mu)|$ &
Rel.\ $\alpha$ limit &
$\log_{10}(\mathrm{Rel.}\ \alpha)$ \\
\hline
Be--Ti & $1.33\times10^{-2}$ & 0.356 & $-0.45$ \\
Be--Al & $3.59\times10^{-2}$ & 0.132 & $-0.88$ \\
Be--W  & $4.29\times10^{-2}$ & 0.111 & $-0.96$ \\
Al--Ti & $2.26\times10^{-2}$ & 0.210 & $-0.68$ \\
Al--W  & $7.89\times10^{-2}$ & 0.0602 & $-1.22$ \\
Ti--W  & $5.63\times10^{-2}$ & 0.0844 & $-1.07$ \\
\end{tabular}
\end{ruledtabular}
\label{tab:neutron_contrast}
\end{table}

The optimal material choice depends strongly on the assumed coupling
model. For baryon coupling, Be--Ti provides the largest contrast among
the material pairs considered here. Be--Al and Be--W give projected
limits within factors of approximately 1.2 and 1.5 of the Be--Ti
baseline, whereas Al--Ti, Al--W, and Ti--W yield weaker limits by
factors of approximately 3--6. Under neutron coupling, the ranking
changes substantially: Al--W provides the largest contrast, followed
by Ti--W and Be--W. Relative to Be--Ti, these pairs improve the
projected neutron-coupling limits by factors of approximately 5.9,
4.2, and 3.2, respectively. Material selection can therefore change
the projected Yukawa sensitivity by factors of several and should be
treated as an important design parameter in future WEP experiments
using differential torsion sensors.

\subsection{Experimental Systematics}
\label{sec:systematics}

With the Earth as the source, a composition-dependent signal is static in the laboratory frame. The modulated projections of Sec.~\ref{sec:results} therefore assume that the signal can be moved to $f_{\rm mod}$, for example by rotating the sensor about its suspension axis, as in rotating torsion-balance experiments, which operate at rotation frequencies of order 1\,mHz~\cite{Schlamminger2008,Wagner_2012}. We take $f_{\rm mod}=2$\,mHz, a turntable period of about 8\,min comparable to the 20-min period of Ref.~\cite{Schlamminger2008}, as the baseline. As a higher-frequency benchmark we take $f_{\rm mod}=20$\,mHz, below the 26\,mHz torsional resonance by a ratio similar to the two-thirds used in Ref.~\cite{Schlamminger2008}; this case assumes a rotation and isolation scheme compatible with a 50-s rotation period. Rotating a multi-stage seismically isolated sensor at 0.2\,Hz would be considerably more demanding, since turntable-induced tilt and other rotation-synchronous disturbances become harder to control at higher rotation rates; the 0.2\,Hz curves therefore indicate the reach of a faster modulation scheme. Sources fixed relative to the Sun or the Galaxy are modulated by the Earth's rotation without a turntable~\cite{Schlamminger2008,Wagner_2012}, but at $\simeq1.2\times10^{-5}$\,Hz, below the band of Fig.~\ref{fig:eta_spectrum}. In addition, the projections in Figs.~\ref{fig:eta_spectrum} and~\ref{fig:alpha_lambda} are noise-limited (SNR\,$=1$), whereas published limits include systematic uncertainties; the MICROSCOPE result~\cite{Touboul2022}, moreover, refers to a Ti--Pt pair rather than Be--Ti.

For scale, the projected $\eta\simeq7.5\times10^{-16}$ at $f_{\rm mod}=2$\,mHz corresponds to a differential acceleration $\Delta a=\eta\,g\,\mathcal{P}\simeq1.2\times10^{-17}$\,m\,s$^{-2}$. In the error budget of Ref.~\cite{Schlamminger2008} for the northward component, the uncertainties of the gravity-gradient, tilt, and temperature-gradient corrections are $0.2\times10^{-15}$, $0.6\times10^{-15}$, and $1.7\times10^{-15}$\,m\,s$^{-2}$, about 17, 50, and 140 times larger. Reaching the projected sensitivity therefore requires a calibrated systematic error budget correspondingly tighter than in existing torsion-balance experiments.

Environmental disturbances and
systematic couplings also need to be monitored and controlled. Magnetic noise
can be reduced by using low-magnetic-permeability materials close to the
torsion pendulums and, if necessary, by enclosing the sensor in
$\mu$-metal shielding. Similar shielding and calibration techniques have
been used in previous torsion-balance WEP experiments to constrain the
residual magnetic coupling~\cite{Schlamminger2008,Wagner_2012}.

In the most sensitive band, Newtonian noise at a surface site can exceed the instrumental noise budget by an order of magnitude or more~\cite{Chua2023,Sun2025}; the projections of Sec.~\ref{sec:results} assume that it is mitigated below the instrumental noise.
Because it cannot be shielded or attenuated by mechanical
isolation, Newtonian noise requires different mitigation techniques. One approach is to
use a second, closely positioned differential torsion sensor with
identical test-mass compositions to measure the local Newtonian
background and subtract its contribution from the WEP sensor. Other
options include operating the experiment in a low-seismic-noise
environment~\cite{Peterson1993.USGS} or underground~\cite{Punturo2010}.
Local density and displacement fluctuations can also be reduced by
replacing soil near the apparatus with lightweight material. In addition,
a seismometer array can be used to measure the local seismic field and
estimate the corresponding Newtonian-noise contribution~\cite{Hall2021}.

Gravity gradients can couple to residual geometrical asymmetries and
multipole moments of the torsion pendulums, producing a differential torque that can mimic or obscure a composition-dependent signal. This coupling can be reduced through symmetric pendulum design.
Previous torsion-balance experiments have also used a dedicated
gradiometer pendulum to measure the relevant gravity-gradient components
and gravity-gradient compensators to reduce their contribution
~\cite{su1994new,Schlamminger2008,Wagner_2012}.

Tilt-to-torque coupling also needs to be controlled. Previous rotating
torsion-balance experiments used tilt sensors and active feedback to
stabilize the rotation axis relative to local vertical. The residual
coupling was measured by deliberately tilting the apparatus and measuring
the resulting feedthrough to the pendulum twist
~\cite{su1994new,heckel2008preferred,Schlamminger2008,Wagner_2012}.

\section{Conclusion}
\label{sec:conclusion}
We have evaluated the projected sensitivity of a differential torsion sensor to violations of the weak equivalence principle and to composition-dependent Yukawa interactions. For a sensor with parameters comparable to an existing prototype~\cite{Chua2023}---5-kg test masses on a 0.24-m lever arm with optical-cavity readout---the projected noise-limited E\"otv\"os sensitivity after 180 days is $\eta\simeq7.5\times10^{-16}$ and $2.4\times10^{-16}$ for signal modulation at 2 and 20\,mHz, below current torsion-balance limits and comparable to the level reached by MICROSCOPE~\cite{Touboul2022}, and $\eta\simeq7.7\times10^{-17}$ for modulation at 0.2\,Hz. With the Earth as the source, this corresponds to projected Yukawa-coupling sensitivities of $\alpha\simeq3.1\times10^{-13}$ and $9.8\times10^{-14}$ for $\lambda\gtrsim10^{7}$\,m.

Beyond the noise floor itself, three ingredients set this reach: the modulation scheme, which fixes the operating frequency for an Earth-fixed source; the material contrast, which changes the reach by factors of several depending on the coupling; and the source geometry, which limits Earth-source projections without a site-specific model to $\lambda\gtrsim10^{7}$\,m. Attaining this reach in practice further requires a systematic error budget tighter than in existing torsion-balance experiments (Sec.~\ref{sec:systematics}).

Several directions could extend this work. Larger test masses would lower the acceleration-equivalent thermal noise, which dominates below $\sim5$\,Hz. A slow turntable compatible with the seismic-isolation platform is a key experimental step toward an Earth-source test. A movable laboratory attractor, with its own source model, or a site-specific topographic model would extend the reach below $10^{7}$\,m. Finally, the same instrument is sensitive to ultralight dark matter~\cite{Sun2025}, so a single sensor could address both static and oscillating composition-dependent forces.
\begin{acknowledgments}
The authors thank Perry Forsyth for providing transfer-function and thermal-noise data of the differential torsion sensor prototype. This
research is supported by the Australian Research
Council Centre of Excellence for Gravitational Wave
Discovery (OzGrav), Projects No. CE170100004 and
No. CE230100016. S. S. Y. C. is supported by the Australian Research Council Discovery Project No. DP250104537. L. S. is also supported by the
Australian Research Council Discovery Early Career
Researcher Award, Project No. DE240100206.
\end{acknowledgments}

\appendix
\section{Yukawa Derivation}
\label{app:yukawa}
Consider a composition-dependent charge-to-mass ratio $\left(q/\mu\right)$, a source mass density $\rho(\mathbf{r})$ carrying $\left(q/\mu\right)_{\rm src}$, and a test body $A$ carrying $\left(q/\mu\right)_{A}$~\cite{Fischbach1998,Adelberger2009}. In accordance with Eq.~\eqref{eq:yukawa-potential}, the Yukawa potential per unit mass of $A$ at position $\mathbf{x}$ is
\begin{equation}
\begin{split}
\Phi_Y^{(A)}(\mathbf{x})
= {}& -\,G\,\alpha\,\!\left(\frac{q}{\mu}\right)_{\!A}
\int \rho(\mathbf{r})\!\left(\frac{q}{\mu}\right)_{\!\rm src}
\frac{e^{-s/\lambda}}{s}\,d^{3}\mathbf{r}, \\
&\qquad s = \lvert \mathbf{x} - \mathbf{r} \rvert .
\end{split}
\end{equation}
\noindent
The acceleration of $A$ is $\mathbf{a}_Y^{(A)}(\mathbf{x})=-\nabla_{\!x}\,\Phi_Y^{(A)}(\mathbf{x})$. Taking the gradient inside the integral gives

\begin{equation}
\begin{split}
\mathbf{a}_Y^{(A)}(\mathbf{x})
= {}& G\,\alpha\,\!\left(\frac{q}{\mu}\right)_{\!A}
\int \rho(\mathbf{r})\!\left(\frac{q}{\mu}\right)_{\!\rm src}
\frac{e^{-s/\lambda}}{s^{2}} \\
&\times \left(1+\frac{s}{\lambda}\right)\hat{\mathbf{s}}\,d^{3}\mathbf{r},
\qquad s=\lvert\mathbf{x}-\mathbf{r}\rvert .
\end{split}
\label{eq:append_acceleration}
\end{equation}
where $\hat{\mathbf{s}}=(\mathbf{r}-\mathbf{x})/s$ is the unit vector from the field point toward the source element.

The composition-dependent signal is the difference between the accelerations of test masses of materials $A$ and $B$, projected onto the sensitive (horizontal) direction of the sensor,

\begin{equation}
\label{eq:delta-a}
\begin{split}
\Delta a_Y(\lambda)&=\,|(\mathbf{a}_Y^{(A)}-\mathbf{a}_Y^{(B)})\,\cdot \hat{\mathbf{u}}_{\rm sens}|\\
&= G\,\alpha\,|\Delta(q/\mu)|\,(q/\mu)_{\rm src}\, S(\lambda)\,\mathcal{P}.
\end{split}
\end{equation}
\noindent
The parameter $\lambda$ denotes the interaction range, equal to the Compton wavelength of the exchanged boson. Physically, $\lambda$ controls the scale of exponential suppression: for $r \ll \lambda$ the force is nearly Coulombic, while for $r \gg \lambda$ the interaction is exponentially screened. Thus $\lambda$ is a property of the new interaction itself, not the source–detector distance.

Due to Earth's rotation, a laboratory-fixed suspension experiences an inertial (centrifugal) acceleration $\vec{a}_c$ in addition to gravity $\vec{g}$. The suspension wire aligns with the net effective acceleration $\vec{g}_{\rm eff} = \vec{g} + \vec{a}_c$, which is slightly tilted from the true vertical. This tilt introduces a projection between the true gravitational direction and the sensitive axis of the torsion sensor. The horizontal component of $\vec{g}$ projected onto the sensor plane is quantified by a small projection factor $\mathcal{P}$, which, at latitude $\varphi$, is approximately
\begin{equation}
\mathcal{P} \simeq \frac{\Omega_\oplus^2 R_\oplus}{g} \lvert\sin\varphi \cos\varphi\rvert.
\label{eq:proj_factor}
\end{equation}
This factor vanishes at the equator and reaches its maximum magnitude near \(45^\circ\). For Canberra, Australia, at latitude \(\varphi=-35.3^\circ\), Eq.~\eqref{eq:proj_factor} gives \(\mathcal{P}\simeq1.63\times10^{-3}\), corresponding to a horizontal acceleration component \(\simeq1.60\times10^{-2}\,\mathrm{m\,s^{-2}}\). This expression is adopted as a long-range approximation, for interaction ranges comparable to or larger than the Earth's radius (see Sec.~\ref{sec:results}).

\section{Noise Derivation}
\label{appx:noise_sources}
For a structural damping model, the fluctuation--dissipation theorem~\cite{Saulson1990} gives the angular thermal-noise spectral density
\begin{equation}
S_\theta^{\rm th}(f)=\frac{4 k_B T}{2\pi f}\,|\Im\!\big[H(f)\big]|,
\label{eq:thermal}
\end{equation}
where $k_B$ is Boltzmann's constant and $T$ is the temperature.

The beam vibration thermal noise is similar to Eq.~\eqref{eq:thermal}. Considering only the fundamental bending mode with effective mass $m_b$, resonant angular frequency $\omega_b$, structural loss angle $\phi_b=Q^{-1}_b$ and mechanical quality factor $Q_b$, the single-mode displacement thermal noise spectral density is

\begin{equation}
    S_x(\omega) =
    \frac{4 k_B T}{m_b\,\omega}
    \frac{\displaystyle \omega_b^2\,\phi_b}
    {\bigl(\omega_b^2 - \omega^2\bigr)^2 + \bigl(\omega_b^2\, \phi_b\bigr)^2 },
    \label{eq:beam_thermal_noise_compact}
\end{equation}
where $\omega=2\pi f$.

Photon shot noise in the optical readout produces fluctuations
in the detected photocurrent, corresponding to a displacement
noise spectral density~\cite{Black2001}

\begin{equation}
    S^{1/2}_{\rm shot}(f) = \frac{\sqrt{hc}}{8}\frac{\sqrt{\lambda_c
    }}{\mathcal{F}\sqrt{P_c}},
\end{equation}
\noindent
where $h$ is Planck's constant, $c$ is the speed of light, $\lambda_c$ is the optical wavelength, $\mathcal{F}$ is the cavity finesse and $P_c$ is the power in the carrier signal.

\section{Source-factor Derivation}
\label{app:source}
\subsection{Exact Source Factor}
For a spherically symmetric source, the Yukawa analogue of the shell theorem states that a thin shell of radius $r'$ and mass $dM$ produces, at an exterior radius $r$, the potential per unit test mass~\cite{Fischbach1998,Fayet2025}
\begin{equation}
\begin{split}
d\Phi_Y(r)={}&-G\alpha\left(\frac{q}{\mu}\right)_{\!A}\left(\frac{q}{\mu}\right)_{\!\rm src}\\
&\times dM\,\frac{\lambda}{r'}\sinh\!\left(\frac{r'}{\lambda}\right)\frac{e^{-r/\lambda}}{r}.
\end{split}
\label{eq:shell}
\end{equation}
Integrating over shells, differentiating with respect to $r$, and setting $r=R_\oplus$ gives the radial differential Yukawa acceleration $\Delta a_Y^{\rm rad}=G\alpha|\Delta(q/\mu)|(q/\mu)_{\rm src}S(\lambda)$, with $S(\lambda)$ given by Eq.~(\ref{eq:S_exact}); its projection onto the sensitive axis is $\Delta a_Y=\Delta a_Y^{\rm rad}\,\mathcal{P}$ [Eq.~(\ref{eq:delta-a})]. Equivalently,
\begin{equation}
S(\lambda)=\frac{M_\oplus}{R_\oplus^{2}}\left(1+\frac{R_\oplus}{\lambda}\right)e^{-R_\oplus/\lambda}\,\Phi\!\left(\frac{R_\oplus}{\lambda}\right),
\end{equation}
where $\Phi(x)=\langle\sinh(xr/R_\oplus)/(xr/R_\oplus)\rangle$ is the mass-weighted hyperbolic form factor of Refs.~\cite{Fayet2025,Fayet2026}. For a stack of homogeneous shells the radial integral is elementary, $\int r\sinh(r/\lambda)\,dr=\lambda^{2}h(r/\lambda)$ with $h(x)=x\cosh x-\sinh x$, so that
\begin{equation}
\begin{split}
S(\lambda)={}&\frac{(1+R_\oplus/\lambda)\,e^{-R_\oplus/\lambda}}{R_\oplus^{2}}\\
&\times\sum_i4\pi\rho_i\lambda^{3}\big[h(R_i^{+}/\lambda)-h(R_i^{-}/\lambda)\big],
\end{split}
\end{equation}
where $R_i^{+}$ and $R_i^{-}$ are the outer and inner radii of shell~$i$. In numerical work the factor $e^{-R_\oplus/\lambda}$ is absorbed into the hyperbolic functions to avoid overflow.

The two limits follow directly. For $\lambda\ll R_\oplus$ only the outermost layer within $\sim\lambda$ of the surface contributes; with $\sinh(r/\lambda)\to e^{r/\lambda}/2$ one finds
\begin{equation}
S_{\rm half}(\lambda)\simeq2\pi\rho_{\rm s}\lambda,
\label{eq:source_half_space}
\end{equation}
in agreement with the half-space calculation of Appendix~\ref{sec:append_derivation_half_space}. For $\lambda\gg R_\oplus$, $\sinh x/x\to1$, $\Phi\to1$, and
\begin{equation}
S_\infty=\frac{M_\oplus}{R_\oplus^{2}},
\label{eq:source_model_large}
\end{equation}
the unscreened Newtonian value (Appendix~\ref{sec:append_asymptotic}). We evaluate Eq.~(\ref{eq:S_exact}) with PREM~\cite{Dziewonski1981}, replacing the 3-km ocean layer by crustal rock ($\rho_{\rm s}=2.6\,\mathrm{g\,cm^{-3}}$) appropriate for a continental site; this model reproduces $M_\oplus$ to $0.1\%$. Using instead the five-shell model of Ref.~\cite{Fayet2025} changes $S(\lambda)$ by less than $4\%$ for $\lambda\geq10$\,km; at shorter ranges $S(\lambda)$ simply scales with the assumed near-surface density.
\subsection{Derivation of the Half-Space Source Factor}
\label{sec:append_derivation_half_space}
We consider a uniform half-space of density~$\rho$ filling $z'<0$, and compute the Yukawa acceleration at a field point $\mathbf{x} = (0,0,z)$ with $z \ge 0$. In cylindrical coordinates about the vertical axis, the downward vertical component, $a_Y^{(z)}\equiv-\hat{\mathbf{z}}\cdot\mathbf{a}_Y^{(A)}$, from Eq.~\eqref{eq:append_acceleration} is

\begin{equation}
\begin{aligned}
a_Y^{(z)}(z,\lambda)
=\, & 2\pi G \alpha
\left( \tfrac{q}{\mu} \right)_{\!A}
\left( \tfrac{q}{\mu} \right)_{\!\rm src} \rho  \int_{-\infty}^{0}\!\! dz' \\
& \times\int_0^\infty \!\! dR\,
\frac{e^{-s/\lambda}}{s^2} \left(1 + \tfrac{s}{\lambda} \right)
\frac{z - z'}{s} \, R,
\end{aligned}
\end{equation}
where $s = \sqrt{R^2 + (z - z')^2}$. Changing variables in the inner integral ($s \leftrightarrow R$) and using
\[
\frac{e^{-s/\lambda}}{s^2} \left(1 + \tfrac{s}{\lambda} \right) = -\frac{d}{ds} \left( \frac{e^{-s/\lambda}}{s} \right),
\]
and evaluating the radial integral, we obtain
\begin{equation}
\int_0^\infty \! dR\;
\frac{e^{-s/\lambda}}{s^2} \left(1 + \tfrac{s}{\lambda} \right)
\frac{z - z'}{s} R
= e^{-(z - z')/\lambda}.
\end{equation}
The remaining integral over $z'$ gives
\[
\int_{-\infty}^{0} e^{-(z - z')/\lambda} dz' = \lambda e^{-z/\lambda},
\]
so the total downward acceleration is
\begin{equation}
a_Y^{(z)}(z,\lambda) =
2\pi G \alpha \left( \tfrac{q}{\mu} \right)_{\!A} \left( \tfrac{q}{\mu} \right)_{\!\rm src}
\rho \lambda e^{-z/\lambda}.
\end{equation}
Comparing with the single-body form $|\mathbf{a}_Y^{(A)}| = G \alpha\, (q/\mu)_A\,(q/\mu)_{\rm src}\, S(\lambda)$, we identify the half-space source factor:
\begin{equation}
S_{\rm half}(z,\lambda) = 2\pi \rho \lambda\, e^{-z/\lambda},
\end{equation}
which reduces at the surface ($z=0$) to the widely used form of Eq.~\eqref{eq:source_half_space}.

\subsection{Asymptotic Source Factor at Large Ranges}
\label{sec:append_asymptotic}

At large Yukawa ranges ($\lambda \to \infty$), the screened interaction kernel becomes unscreened:
\[
\frac{e^{-s/\lambda}}{s^2}\left(1+\frac{s}{\lambda}\right)
\;\longrightarrow\;
\frac{1}{s^2},
\]
recovering the Newtonian field of the source mass distribution.

For a spherically symmetric Earth and a detector at the surface ($|\mathbf{x}| = R_\oplus$),
the shell theorem (or Gauss's law for gravity) gives:
\[
\int \rho(\mathbf{r})\,\frac{\hat{\mathbf{s}}}{s^2}\,d^3\mathbf{r}
\;=\; -\frac{M_\oplus}{R_\oplus^2}\,\hat{\mathbf{r}},
\]
where $\hat{\mathbf{r}}$ is the outward radial unit vector at the detector, so the source factor saturates to Eq.~\eqref{eq:source_model_large}.

\bibliography{refs}

\end{document}